\documentclass[aps,prl,twocolumn,groupedaddress,showpacs,showkeys]{revtex4-1}

\usepackage{graphicx}
\usepackage{float}
\usepackage{amsmath}
\usepackage{xfrac}
\usepackage{upgreek}
\usepackage{bm}
\usepackage{caption}
\usepackage{ragged2e}
\usepackage[version=3]{mhchem}

\renewcommand{\raggedright}{\leftskip=0pt \rightskip=0pt plus 0cm}

\begin{document}

\renewcommand{\thefootnote}{\fnsymbol{footnote}}
\title{Magnon Valves Based on YIG/NiO/YIG All-Insulating Magnon Junctions}

\author{C. Y. Guo}
\thanks{These two authors contributed equally to this work.}
\affiliation{Beijing National Laboratory for Condensed Matter Physics, Institute of Physics, University of Chinese Academy of Sciences, Chinese Academy of Sciences, Beijing 100190, China.}

\author{C. H. Wan}
\thanks{These two authors contributed equally to this work.}
\affiliation{Beijing National Laboratory for Condensed Matter Physics, Institute of Physics, University of Chinese Academy of Sciences, Chinese Academy of Sciences, Beijing 100190, China.}

\author{X. Wang}
\affiliation{Beijing National Laboratory for Condensed Matter Physics, Institute of Physics, University of Chinese Academy of Sciences, Chinese Academy of Sciences, Beijing 100190, China.}

\author{C. Fang}
\affiliation{Beijing National Laboratory for Condensed Matter Physics, Institute of Physics, University of Chinese Academy of Sciences, Chinese Academy of Sciences, Beijing 100190, China.}

\author{P. Tang}
\affiliation{Beijing National Laboratory for Condensed Matter Physics, Institute of Physics, University of Chinese Academy of Sciences, Chinese Academy of Sciences, Beijing 100190, China.}

\author{W. J. Kong}
\affiliation{Beijing National Laboratory for Condensed Matter Physics, Institute of Physics, University of Chinese Academy of Sciences, Chinese Academy of Sciences, Beijing 100190, China.}

\author{M. K. Zhao}
\affiliation{Beijing National Laboratory for Condensed Matter Physics, Institute of Physics, University of Chinese Academy of Sciences, Chinese Academy of Sciences, Beijing 100190, China.}

\author{L. N. Jiang}
\affiliation{Beijing National Laboratory for Condensed Matter Physics, Institute of Physics, University of Chinese Academy of Sciences, Chinese Academy of Sciences, Beijing 100190, China.}

\author{B. S. Tao}
\affiliation{Beijing National Laboratory for Condensed Matter Physics, Institute of Physics, University of Chinese Academy of Sciences, Chinese Academy of Sciences, Beijing 100190, China.}

\author{G. Q. Yu}
\affiliation{Beijing National Laboratory for Condensed Matter Physics, Institute of Physics, University of Chinese Academy of Sciences, Chinese Academy of Sciences, Beijing 100190, China.}

\author{X. F. Han}
\email{Corresponding Author: xfhan@iphy.ac.cn}
\affiliation{Beijing National Laboratory for Condensed Matter Physics, Institute of Physics, University of Chinese Academy of Sciences, Chinese Academy of Sciences, Beijing 100190, China.}


\date{\today}

\begin{abstract}
As an alternative angular momentum carrier, magnons or spin waves can be utilized to encode information and breed magnon-based circuits with ultralow power consumption and non-Boolean data processing capability. In order to construct such a circuit, it is indispensable to design some electronic components with both long magnon decay and coherence length and effective control over magnon transport. Here we show that an all-insulating magnon junctions composed by a magnetic insulator (MI$_1$)/antiferromagnetic insulator (AFI)/magnetic insulator (MI$_2$) sandwich (Y$_3$Fe$_5$O$_{12}$/NiO/Y$_3$Fe$_5$O$_{12}$) can completely turn a thermogradient-induced magnon current on or off as the two Y$_3$Fe$_5$O$_{12}$ layers are aligned parallel or anti-parallel. The magnon decay length in NiO is about 3.5$\sim$4.5 nm between 100 K and 200 K for thermally activated magnons. The insulating magnon valve (magnon junction), as a basic building block, possibly shed light on the naissance of efficient magnon-based circuits, including non-Boolean logic, memory, diode, transistors, magnon waveguide and switches with sizable on-off ratios.
\end{abstract}
\maketitle


\section{Introduction}
Data processing and transmission in sophisticated microelectronics rely strongly on electric current, which inevitably wastes a large amount of energy due to Joule heating. Magnons represent the collective excitations in magnetic systems. Though charge neutral, they possess angular momenta and can also transfer the momenta as information carrier free from Joule heating~\cite{ref1}. The main dissipation channel of magnons in magnetic insulators is spin-lattice coupling which is much weaker than Joule heating. Moreover, the wave nature of magnons provides additional merits, (1) long propagation distance up to millimeters~\cite{ref2,ref3} and (2) a new degree of freedom (magnon phase) owing to which non-Boolean logic processing~\cite{ref1,ref4,ref5,ref6} are anticipated. These characteristics make magnons the ideal information carriers based on which some electronic components for future magnonic circuits are being developed recently~\cite{ref7,ref8,ref9,ref10,ref11}.

Bender \emph{et al.}~\cite{ref7} theoretically proposed a spin valve with magnetic insulator (MI)/nonmagnetic metal/MI structure where magnetization switching induced by thermally driven spin torques was expected. Wu \emph{et al.}~\cite{ref8} proposed a concept of magnon valve with heterostructure of MI/spacer/MI and experimentally realized the heterostructure in a YIG/Au/YIG sandwich whose spin Seebeck effect (SSE) depends on the relative orientation of the top and bottom YIG layers. In this structure, magnon transport occurs in two YIG layers while spin transport in Au remains limited to the electrons. Angular momentum transfer through the structure thus relies on mutual conversion between magnon spin current and electron spin current. Cramer \emph{et al.}~\cite{ref9} prepared another type of hybrid spin valves in YIG/CoO/Co. In this valve, inverse spin Hall voltage (ISHE) in Co induced by spin pumping effect is determined by spin configurations between YIG and Co. The ferromagnetic Co electrode can support both electron and magnon currents, and thus the output signal have both electron spin and magnon contributions ~\cite{ref3,ref9}. In the above cases, additional conversions between magnon current and spin current or vice visa could first reduce effective decay length of angular momentum. Second, more noticeably, magnon phase would be lost in the above conversions since phase information cannot be encoded by ordinary spin current.

Thus, it is very desirable to construct pure magnon valves in an all-insulating structure such that the spin information propagation is uniquely limited to magnons. Very recently, a sandwich consisting of two ferromagnetic insulators and an antiferromagnetic spacer was proposed by Cheng \emph{et al.}~\cite{ref11} where both giant spin Seebeck effect and magnon transfer torques were predicted.

Here we design and further experimentally realize a typical MI/antiferromagnetic insulator (AFMI)/ MI heterostructure using YIG/NiO/YIG sandwiches. We entitle such a MI/AFMI/MI heterostructure as insulating magnon junction (IMJ) for short. Output magnon current of an IMJ generated by SSE can be regulated by its parallel (P) or antiparallel (AP) states. Especially, the output magnon current can be totally shut down in the AP state while superposed in the P state near room temperature, contributing to a large on-off ratio. Demonstration of the pure magnon junction based on all-insulators may further help to develop magnon-based circuits with fast speed and ultralow energy dissipation in the coming future.

\begin{figure*}[thb!]
\captionsetup{labelformat=default,labelsep=space} 
\includegraphics[width=12cm]{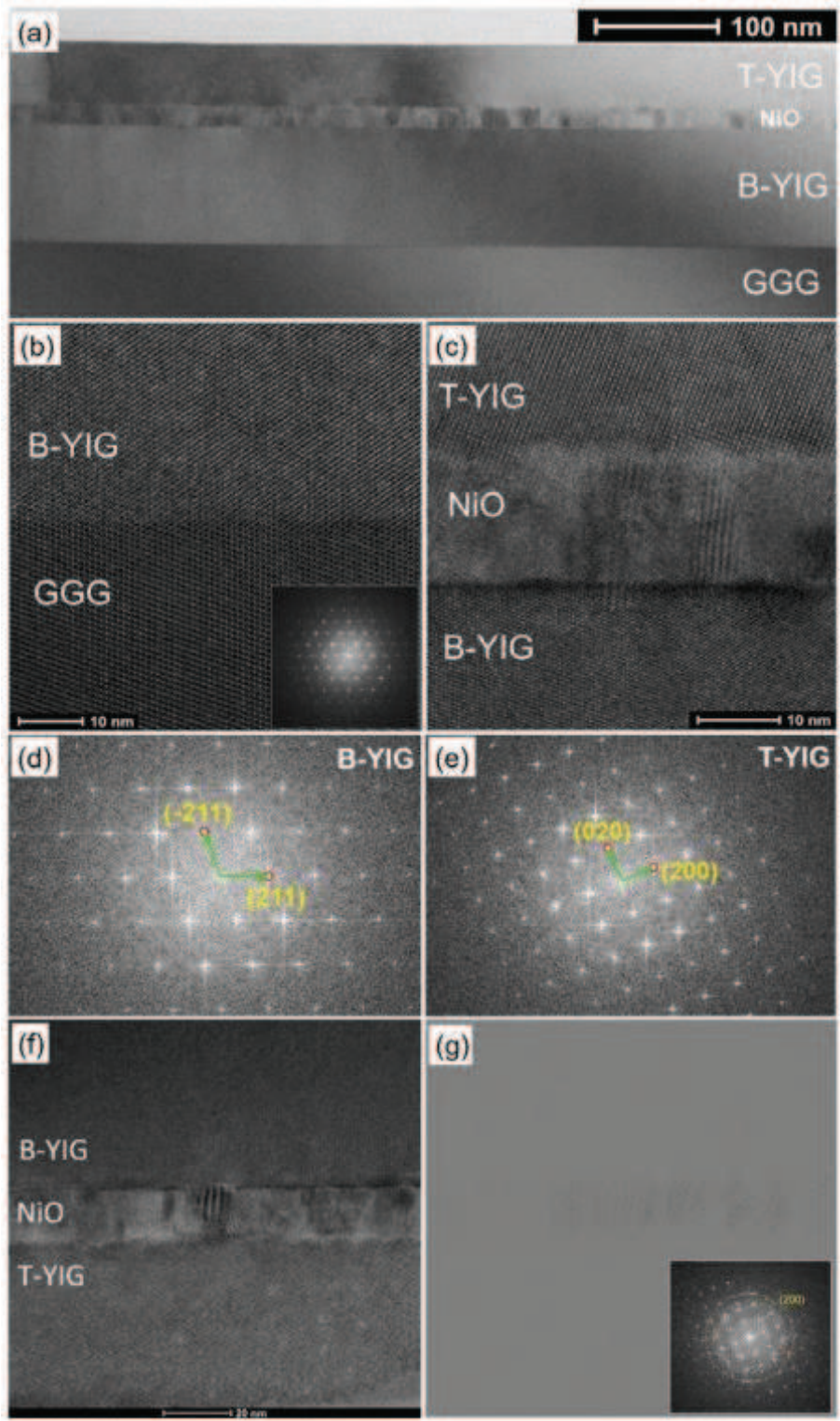}
\caption{\label{Fig1} \raggedright  Microstructure of the GGG//YIG(100)/NiO(15)/YIG(60 nm) IMJ. (a) The cross-sectional TEM image of the sample. HRTEM images of (b) the GGG//YIG(100 nm) and (c) YIG(100)/NiO(15)/YIG(60 nm) interfaces, respectively. Fourier transformation of HRTEM for (d) the bottom YIG and (e) the top YIG only. (f) Larger area HRTEM for the YIG/NiO(15 nm)/YIG IMJ and (g) inverse Fourier Transformation of the yellow ring in the inset diffraction pattern which is obtained by Fourier transformation of Fig.1(f).}
\end{figure*}

\section{Experiments}
IMJs stacks YIG(100)/NiO($t$)/YIG(60 nm) ($t$ = 4, 6, 8, 10, 15, 20, 30, 60, all thickness number in nanometers) were deposited on Gd$_3$Ga$_5$O$_{12}$ (GGG) (111) substrates in a sputtering system (ULVAC-MPS-4000-HC7 model) with base vacuum of 1$\times$10$^{-6}$ Pa. After deposition, high temperature annealing in an oxygen atmosphere was carried out to further improve the crystalline quality of the YIG layers. The stacks with $t$ = 6, 8, 15, 20, 30 and 60 nm were fabricated in the same round. Then a 10 nm Pt stripe with 100 $\mu$m$\times$1000 $\mu$m lateral dimensions for SSE measurement were fabricated by standard photolithography combined with an argon-ion dry etching process. Finally, an insulating SiO$_2$ layer of 100 nm and a Pt/Au stripe were successively deposited on top of the Pt stripe for on-chip heating. Before platinum deposition, vibrating sample magnetometer (VSM, EZ-9 from MicroSense) was used to characterize magnetic properties of the YIG layers. After microfabrication process, SSE of the IMJs were measured in a physical property measurement system (PPMS-9T from Quantum Design). Keithley 2400 provided a heating current ($I$) to the Pt/Au stripe while Keithley 2182 picked up a voltage along the Pt stripe induced by SSE and ISHE. Magnetic field was applied along the transverse direction of the Pt stripe. Control samples YIG(100)/Pt(10 nm), YIG(100)/NiO(15)/Pt(10 nm) and NiO(15)/YIG(60)/Pt(10 nm) were also prepared on GGG substrates and measured for comparison. We have also confirmed insulating nature of oxide parts in the YIG/NiO/YIG and the control stacks by electrical transport measurements.

\section{Results and Discussion}
\section{(1) Structure Characterization}
Fig.1 shows crystalline structure of an IMJ stack GGG//YIG(100)/NiO(15)/YIG(60 nm). The NiO spacer has uniform thickness without pinholes  (Fig.1(a)). The bottom YIG (B-YIG) is epitaxially grown on GGG substrate with atomically sharp interface (Fig.1(b)). Fourier transformation of the high-resolution transmission electron microscope (HRTEM) in the inset only shows one diffraction pattern, further confirming the epitaxial relation. However, the interfaces of the NiO spacer with adjacent YIG layers become rougher than the GGG//YIG interface, especially for the interface with the top YIG (T-YIG) (Fig.1(c)). It is worth noting the two YIG layers are both single-crystalline but with different orientations (Fig.1(d) and (e)). The B-YIG grows along [111] direction of the substrate while the T-YIG does not, probably because the polycrystalline NiO spacer broke epitaxial relation (Fig.1(c)). Fig.1(f) shows a HRTEM image acquired across the bottom-YIG/NiO/top-YIG interfaces. Two sets of diffraction patterns corresponding to Fig.1(d) and Fig.1(e) can be identified. From the patterns and lattice parameters of YIG, we can accurately calibrate camera length of the TEM. Then, besides of the already-known patterns owing to YIGs, diffraction patterns owing to NiO can be also identified as highlighted by the yellow ring and the two red circles in Fig.1(g) inset. The yellow ring corresponds to (200) plane of NiO while the red circles correspond to (111) plane of NiO. After inversely Fourier transforming the yellow ring pattern, NiO polycrystals can be clearly observed as shown in Fig.1(g).

\section{(2) Spin Seebeck Effect Measurement}
Fig.2(a) schematically shows setup to measure SSE of an IMJ. A Pt/Au electrode on top of a 100 nm SiO$_2$ insulating layer is heated by a current and then a temperature gradient $\nabla T$ along the stack normal (+$z$ axis) is built. $\nabla T$ introduces inhomogeneous distribution of magnons inside a MI and produce a magnon current along $\nabla T$ ~\cite{ref12,ref13}. The magnon current can be further transformed as a spin current penetrating into an adjacent heavy metal and then generate a sizable voltage by ISHE~\cite{ref14,ref15,ref16,ref17}, which is so-called longitudinal SSE. Here we use a Pt stripe to measure the voltage ($V_\mathrm{SSE}$) induced by ISHE and monitor the magnitude and direction of spin current exuded from the top YIG. Fig.2(b) shows angle scanning of $V_\mathrm{SSE}$ of an IMJ with $t_\mathrm{NiO}$=8 nm. The Pt stripe is along the $y$ axis. Therefore SSE can be observed only if magnetization has component in the $x$ axis. This requirement gives the angle dependence shown in Fig.2(b). Fig.2(c) shows spin Seebeck voltage as a function of applied field measured at different heating currents. Saturated $V_\mathrm{SSE}$ parabolicly depends on current (Fig.2(d)). This parabolic dependence confirms the thermopower essence of the measured voltage signals while the angle dependences confirm the voltage signals are induced by spin Seebeck effect.
We have also estimated temperature rise ($\Delta T$) of spin detector (Pt stripe) as elevating heating current by calibrating resistivity of the Pt stripe. Heating current of 10 mA, 15 mA and 20 mA would lead to $\Delta T$ of 2.7 K, 6.0 K and 10 K, respectively, as background temperature within (50 K, 325 K). In order to enhance signal-to-noise ratio and minimize influence of heating current on IMJs, we have selected 15 mA as heating current to conduct the following SSE measurement in Fig.4 and Fig.5.

\begin{figure}[thb!]
\captionsetup{labelformat=default,labelsep=space} 
\includegraphics[width=9cm]{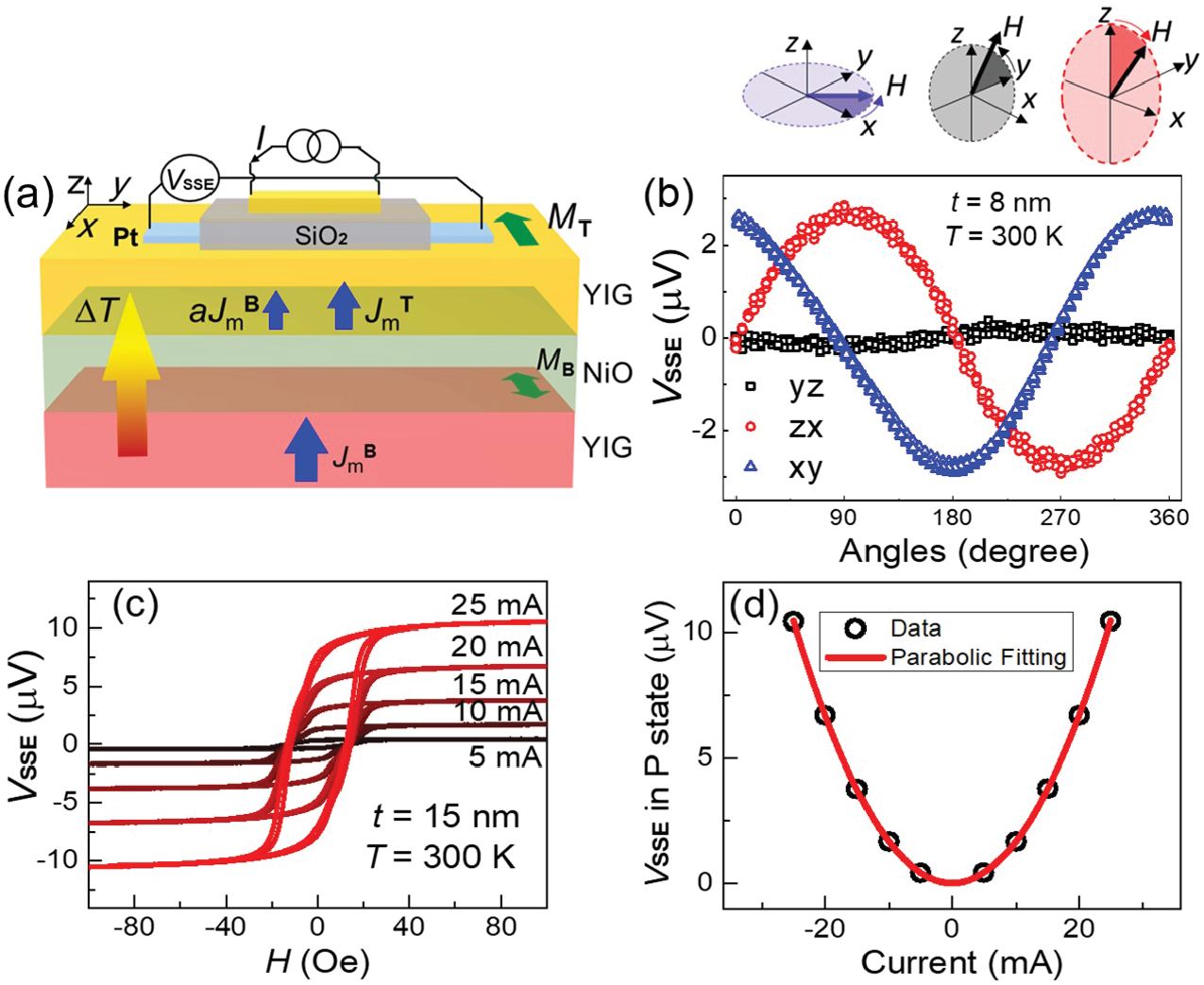}
\caption{\label{Fig2} \raggedright (a) Schematic diagram of spin Seebeck effect and its measurement setup for an IMJ. During measurement, a field in the $x$-axis is applied. (b) Angle scanning of $V_\mathrm{SSE}$ of an IMJ with $t_\mathrm{NiO}$=8 nm. The Pt stripe is along the $y$ axis. (c) Spin Seebeck voltage as a function of applied field measured at elevated heating currents. (d) Parabolic fitting of the field dependence of saturated $V_\mathrm{SSE}$.}
\end{figure}

\section{(3) Field Dependence of \textbf{$V_\mathrm{SSE}$} of an Insulating Magnon Junction}
Fig.3(a) shows a typical hysteresis of an IMJ YIG(100)/NiO(8)/YIG(60 nm). Two magnetization reversals have been identified. Epitaxial growth and larger thickness endow the bottom YIG with lower coercivity ($H_\mathrm{C}$) and larger magnetization than the top YIG on the NiO spacer. Thus, the reversal with $H_\mathrm{C}\approx$ 3 Oe and $\Delta M\approx$ 1.3¡Á($M_\mathrm{S,B}+M_\mathrm{S,T}$) is attributed to the switching of the bottom YIG while the other reversal with $H_\mathrm{C}\approx$ 16 Oe and $\Delta M\approx$ 0.7¡Á($M_\mathrm{S,B}+M_\mathrm{S,T}$) is attributed to the top YIG. $M_\mathrm{S,B}$ and $M_\mathrm{S,T}$ are saturated magnetization of the bottom and top YIG layers, respectively. The obtained $M_\mathrm{S,B}$/$M_\mathrm{S,T}$ is 13/7, close to the ratio 5/3 in the nominal thickness. Due to the difference in coercivity, parallel and antiparallel spin configurations can be formed, as illustrated in the figure.

Fig.3(b) shows field-dependence of $V_\mathrm{SSE}$. Besides of a large $\Delta V_\mathrm{SSE} \approx$ 1.6 $V_\mathrm{SSEmax}$ occurring at 13 Oe, $V_\mathrm{SSE}$ also sharply changes by about 0.4$V_\mathrm{SSEmax}$ at 2 Oe. $V_\mathrm{SSEmax}$ is the saturation value (2.6 $\mu$V) in Fig.3(b). d$V_\mathrm{SSE}$/d$H$ and d$M$/d$H$ (Fig.3(c)) are used to show correspondence between SSE and VSM results. A peak in Fig.3(c) represents a sharp reversal of a YIG layer. There are four peaks in both field dependences. The middle two labeled as (P1-) and (P1+) originate from the reversal of the bottom YIG while the outer ones marked as (P2-) and (P2+) are caused by the reversal of the top YIG. SSE of the control samples YIG/NiO/Pt, NiO/YIG/Pt and YIG/Pt have also been measured (Fig.3(d)). Except YIG/NiO/Pt, the other samples show comparable $V_\mathrm{SSE}$ at the same heating current, which may be owing to higher spin mixing conductance of YIG/Pt interface than that of NiO/Pt interface in our case. YIG/NiO/Pt and YIG/Pt are indeed much softer than NiO/YIG/Pt. Furthermore, only one magnetization reversal is observed for the control samples. If the NiO spacer is replaced by an MgO spacer, $V_\mathrm{SSE}$ signal due to the reversal of the bottom layer disappears as shown in Ref~\cite{ref8}. The above observation indicates that the magnon current from the bottom layer can flow through the NiO spacer and the top YIG and finally penetrate into Pt. Magnon decay length in epitaxial and polycrystalline YIG is about 10 $\upmu$m~\cite{ref2} and several tens of nanometers~\cite{ref18,ref19}, respectively. Wang \emph{et al}.~\cite{ref20} reported spin relaxation length of 9.8 nm in NiO. Our IMJs have comparable dimensions with those reported values, indicating magnon current from the bottom YIG layer capable of flowing into Pt.

Remarkably, though solidly confirmed in experiments, longitudinal spin Seebeck effect is regarded to be potentially caused by (1) difference in electron temperature and magnon temperature across an interface between heavy metal and magnetic insulator~\cite{ref12,ref13} or by (2) inhomogeneous magnon distribution inside bulk region of a magnetic insulator and as-induced pure magnon flow~\cite{ref21,ref22}. These two mechanisms are hard to tell in classic magnetic insulator/heavy metal bilayer systems. Though not ruled out possibility of the 1$^\mathrm{st}$ mechanism, nevertheless, our experiment strongly proved rationality of the 2$^\mathrm{nd}$ mechanism since the bottom YIG layer could only deliver magnon current into Pt via the bulk effect.

\begin{figure}[thb!]
\captionsetup{labelformat=default,labelsep=space} 
\includegraphics[width=9cm]{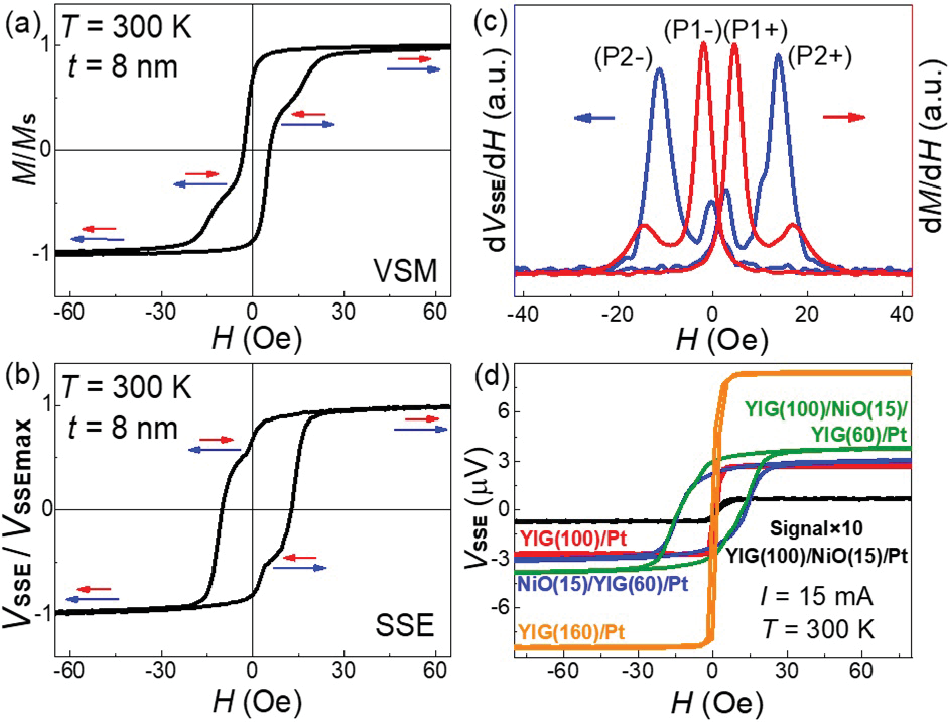}
\caption{\label{Fig3} \raggedright (a) Hysteresis loop and (b) field dependence of $V_\mathrm{SSE}$ for an IMJ with $t$ = 8 nm at 300 K. (c) Corresponding field dependences of d$V_\mathrm{SSE}$/d$H$ and d$M$/d$H$. (d) $V_\mathrm{SSE}$ of the control samples.}
\end{figure}

\section{(4) \textbf{$T$} and $t_\mathrm{NiO}$ Dependence of SSE of Insulating Magnon Junctions}
We have measured field dependence of $V_\mathrm{SSE}$ at 15 mA with elevating $T$ for different IMJs as shown in Fig.4. In order to distinguish switching fields from different YIG layers, their d$V_\mathrm{SSE}$/d$H$ are shown in Fig.5. First, all IMJs show a significant exchange bias below blocking temperature of about 100 K (Fig.4, Fig.5 and Fig.6(b)), which evidences the appearance of NiO antiferromagnetism. Similar blocking temperatures for all the samples indicate interfacial nature of exchange bias effect whose magnitude is dominantly determined by exchange coupling strength between interfacial layers of NiO and YIG.

\begin{figure*}[thb!]
\captionsetup{labelformat=default,labelsep=space} 
\includegraphics[width=17cm]{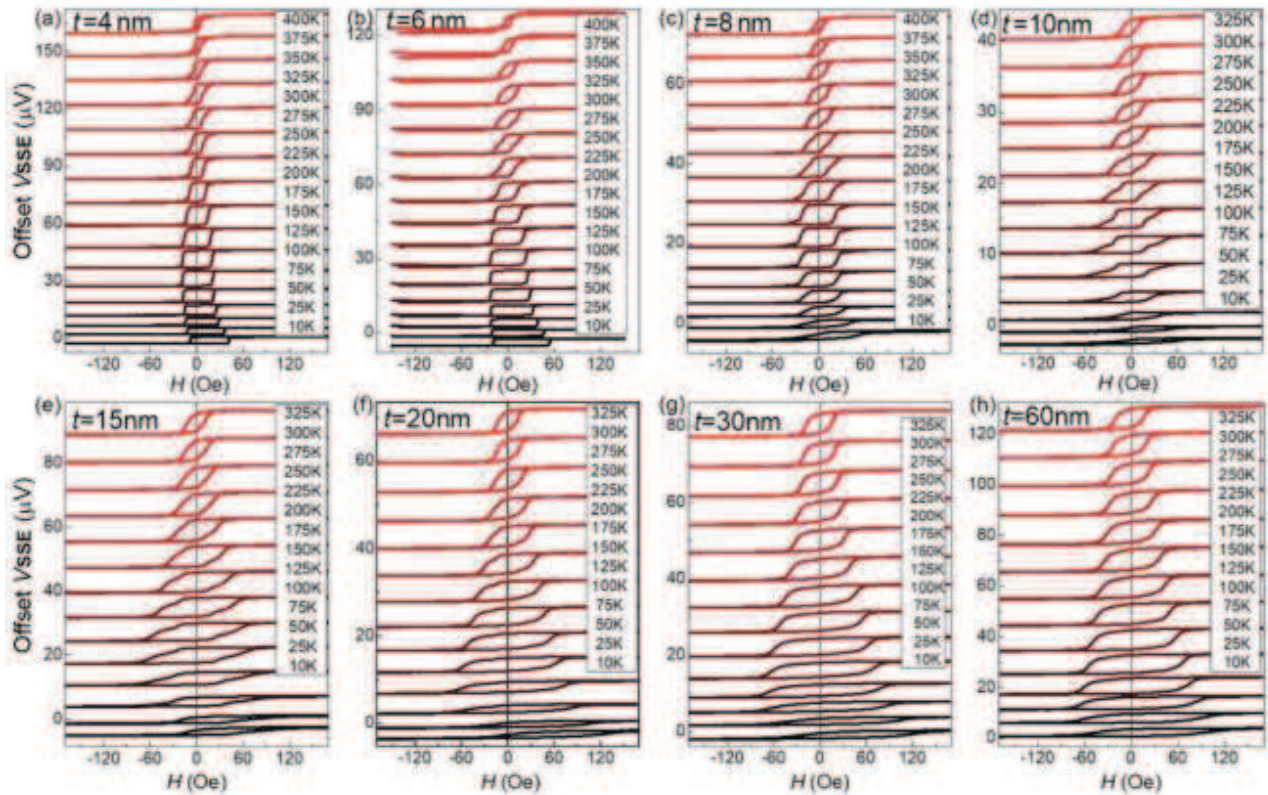}
\caption{\label{Fig2} \raggedright Field dependence of $V_\mathrm{SSE}$ at elevated temperatures for IMJs with (a-h) $t$=4 nm, 6 nm, 8 nm, 10 nm, 15 nm, 20 nm, 30 nm and 60 nm, respectively.}
\end{figure*}

Second, only two peaks are unambiguously identifiable for the IMJ with 30 nm and 60 nm NiO at all temperatures (Fig.5(g,h)). These peaks belong to (P2+) and (P2-) because their positions are identical with those of YIG(100)/NiO(15)/YIG(60 nm) determined by VSM and $H_\mathrm{C}$ of NiO/YIG/Pt as shown in Fig.6(a). Due to too thick NiO spacer and its blocking effect on magnon current, the Pt stripe can only detect magnon current from the top YIG. Thus it is nature that the peaks in Fig.5(g,h) share the same positions with P2+ and P2-.

For $t$=6$\sim$20 nm, temperature evolution of d$V_\mathrm{SSE}$/d$H$ vs. $H$ curves seem nontrivial. Within a certain temperature region, 4 peaks corresponding to 4 switching events of two YIG layers can be clearly resolved. For the $t$ = 15 nm IMJ, for example, four peaks can be clearly resolved between 50 K and 275 K. The relative intensity of (P1+/P2+) or (P1-/P2-) decreases gradually with increasing $T$ (Fig.5(e)). The trend has also been reproduced in the IMJ with $t$ = 8 nm (Fig.5(c)). At $T$  $<$ 175 K, the pair of peaks (P1+) and (P1-) are dominant. At intermedium temperature from 175 K to 325 K, the other pair of peaks (P2+) and (P2-) emerge and are enhanced with increasing $T$. At $T$ $>$ 325 K, (P1+) and (P1-) eventually fade away but (P2+) and (P2-) remain (Fig.5(c)). This trend indicates the layer dominating SSE of an IMJ can be changed between the two YIG layers though spin detector is only directly connected to the top layer.

The IMJ with 4 nm NiO is unique, in which only P1+ and P1- are observed at all temperatures, which is also indicated in Fig.6(a). Though P2+ and P2- are absent here, two broad shoulders outsides of P1+ and P1- can be identified above 275 K, which probably still originate from switching of the top YIG. For thin enough NiO spacer, magnon current generated in the bottom layer can still survive after a weak decay in the NiO spacer at high temperatures. Thus $V_\mathrm{SSE}$ due to the bottom layer is still observable or even dominated in this case.

\begin{figure*}[thb!]
\captionsetup{labelformat=default,labelsep=space} 
\includegraphics[width=17cm]{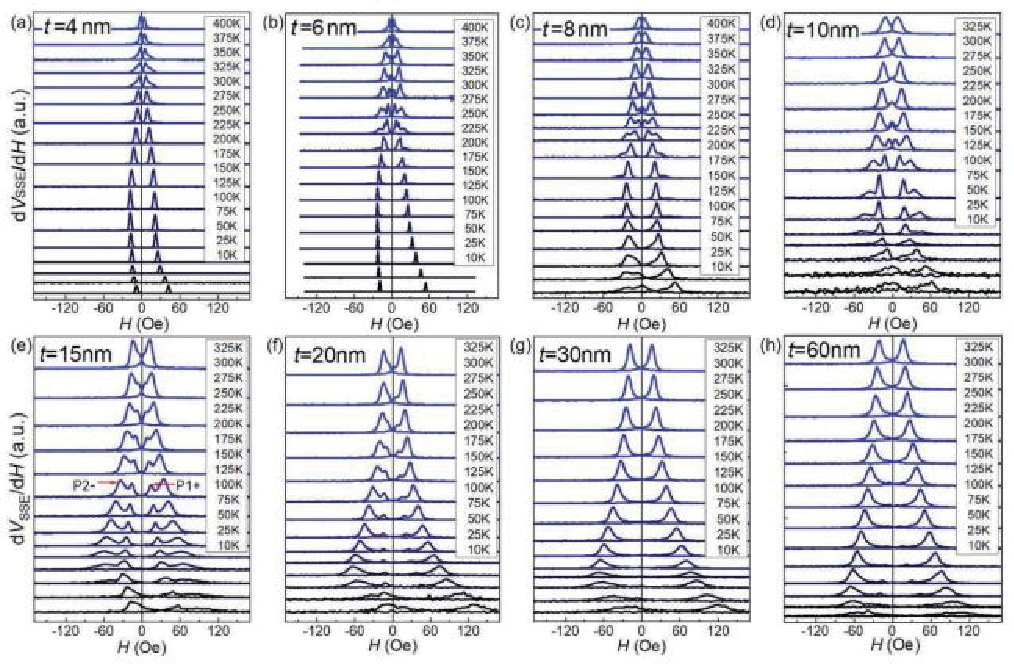}
\caption{\label{Fig3} \raggedright Field dependence of d$V_\mathrm{SSE}$/d$H$  at elevated temperatures for IMJs with (a-h) $t$=4 nm, 6 nm, 8 nm, 10 nm, 15 nm, 20 nm, 30 nm and 60 nm, respectively.}
\end{figure*}

\section{(5) Magnon decay length of NiO}
In order to further check the correlation between the peaks at different temperatures in Fig.5 and switching fields of the YIG layers, we have plotted them (open symbols) together with $H_\mathrm{C}$ of YIG/NiO/Pt and NiO/YIG/Pt (solid triangles) determined by SSE and $H_\mathrm{C}$ of YIG/NiO(15)/YIG determined by VSM (solid hexagons) in Fig.6(a). Remarkably, nearly all the peaks lie on 4 branches defined by $H_\mathrm{C}$ of the control samples and $H_\mathrm{C}$ from the VSM results, which unambiguously demonstrates origin of the four peaks in the entire temperature range, i.e., the outer peaks from the top and the inner peaks from the bottom YIG. Thus, we can conclude the magnon current that overwhelmingly contributes to SSE comes from the bottom YIG at low temperatures while the magnon current from the top YIG becomes significant at high temperatures. A plausible explanation of the relative contributions of the magnon current from the two YIG layers is as follows. The bottom YIG grown on GGG has much better quality and thus a larger SSE coefficient. At low temperature and for a small NiO thickness, the magnon current from the bottom layer is able to propagate through both NiO and the top YIG without noticeable decay. When NiO becomes thicker, magnon current from the bottom YIG decays significant. On the other hand, the magnon current of the top YIG does not suffer such decay since it is in direct contact with Pt. Thus the bottom and top YIG contribute more dominantly at low and high temperatures, respectively.

Fig.6(c) summarizes magnon valve ratio $\eta_\mathrm{mv}$ of the IMJs as a function of $T$. Here $\eta_\mathrm{mv}$ is defined as $ V_\mathrm{SSE,AP}/V_\mathrm{SSE,P}$. $V_\mathrm{SSE}$ induced by ISHE is proportional to the injected magnon current $J_\mathrm{m}$ flowing toward Pt from the top YIG layer. According to the SSE theory~\cite{ref12,ref13,ref21,ref22}, $J_\mathrm{m,T/B}$ is proportional to $S_\mathrm{T/B}(\nabla T)_\mathrm{T/B}$. Here $J_\mathrm{m,T}$ is the spin Seebeck coefficient of the top/bottom YIG and $(\nabla T)_\mathrm{T/B}$ is the induced temperature gradient across the top/bottom YIG and proportional to $I^2$. Thus $V_\mathrm{SSE} \propto I^2$, as confirmed in Fig.2(d). The $J_\mathrm{m,P/AP}=J_\mathrm{m,T} \pm a_\mathrm{NiO} a_\mathrm{YIG} J_\mathrm{m,B}$ in the P and AP states. Here, $J_\mathrm{m,T}$ and $J_\mathrm{m,B}$ are the generated magnon currents by SSE in the top and bottom YIG, respectively, while $a_\mathrm{NiO}$ and $a_\mathrm{YIG}$ are the decay ratio of the magnon current from the bottom YIG in NiO and the top YIG, respectively. The magnon valve ratio $\eta_\mathrm{mv}$ can be thus rewritten as

\begin{equation}
\eta_\mathrm{mv}=\frac{J_\mathrm{m,AP}}{J_\mathrm{m,P}}=\frac{J_\mathrm{m,T} - a_\mathrm{NiO} a_\mathrm{YIG} J_\mathrm{m,B}}{J_\mathrm{m,T} + a_\mathrm{NiO} a_\mathrm{YIG} J_\mathrm{m,B}}
\end{equation}

$\eta_\mathrm{mv}$ sign reflects the relative magnitude of the magnon currents from the top and the bottom layers. The negative values in Fig.6(c) indicates the magnon current from the bottom YIG is larger at low temperature. The positive $\eta_\mathrm{mv}$ seen in the 6 and 8 nm NiO IMJs of Fig.6(c) at high temperatures means a larger magnon current from the top YIG. The most interesting case is $\eta_\mathrm{mv}$ = 0 where the net magnon current at the YIG/Pt interface becomes zero, i.e., the exact cancellation of the two magnon currents generated by two YIG layers (inset of Fig.6(c)). Such cancellation only occurs at the AP state. The figure also shows a trend that the critical temperature where $\eta_\mathrm{mv}$=0 increases with decreasing $t_\mathrm{NiO}$.

Next, we estimate the magnon decay length $\lambda_\mathrm{NiO}$ in the NiO spacers. Magnon decay ratio in NiO is $a_\mathrm{NiO}$. From Eq.(1), one can easily see $a_\mathrm{NiO}$ is proportional to $\delta$ = (1-$\eta_\mathrm{mv}$)/(1+$\eta_\mathrm{mv}$). We then plot ln$\delta$ as a function of $t_\mathrm{NiO}$ between 100 K to 200 K (Fig.6(d)). In this temperature region, the switching fields of both YIGs are well separated for the IMJs. Worth noting, data from the IMJs with 4 nm and 10 nm NiO spacer are not used here. For the IMJ with 4 nm NiO, only two peaks are clearly observed (Fig.5(a)). Thus it is hard to obtain reliable $\eta_\mathrm{mv}$ for the device. For the IMJs with 10 nm and 4 nm NiO, the stacks were deposited and annealed in different rounds with the others. The other stacks were fabricated in the same round and their data were systematic and thus more comparable. Fig.6(d) shows a good linear dependence of ln$\delta$ on $t_\mathrm{NiO}$, which suggests that magnons generated in the bottom YIG can pass through NiO in a diffusive way~\cite{ref11}. The inset in Fig.6(d) shows $T$-dependence of the derived $\lambda_\mathrm{NiO}$ which increases slightly with $T$. $\lambda_\mathrm{NiO}$ is about 3.5 nm $\sim$ 4.5 nm between 100 K and 200 K. Though $\lambda_\mathrm{NiO}$ slightly increases with $T$. However, $V_\mathrm{SSE}$ from the top YIG gradually dominates at higher temperatures. It indicates not only magnon transport properties of an IMJ but also $T$-dependent spin Seebeck coefficients of YIG layers and thermoconductivity of YIG and NiO layers will eventually affect the magnon valve effect. The influence of these parameters is already out of scope of this article. Therefore, we only use magnon valve ratio and its thickness dependence at a fixed temperature to measure magnon decay length as shown in the main panel of Fig.6(d).

Spin decay length in antiferromagnetic materials such as NiO, CoO, Cr$_2$O$_3$ and IrMn have been measured by spin pumping or spin Hall magnetoresistance techniques~\cite{ref20,ref23,ref24,ref25,ref26,ref27,ref28,ref29}. For NiO thin films, $\lambda_\mathrm{NiO}$ is reported about 10 nm~\cite{ref20,ref23}. Our value has the same order of magnitude with theirs. Qiu $et$  $al$ ~\cite{ref27,ref28} reported an enhanced spin pumping effect near $\rm N\acute{e}el$ temperature $T_\mathrm{N}$ of antiferromagnetic materials. For 6 nm CoO and 1.5 nm NiO, they observed the most significant enhancement at about 200 K and 285 K, respectively~\cite{ref27}. In our case, NiO spacers have thickness of 4 nm $\sim$ 60 nm. They probably have even higher $T_\mathrm{N}$. Then the increase in $\lambda_\mathrm{NiO}$ within (100 K, 200 K), we think, is also due to similar enhancement in magnon transport efficiency as $T$ approaching $T_\mathrm{N}$.

\begin{figure}[thb!]
\captionsetup{labelformat=default,labelsep=space} 
\includegraphics[width=9.5cm]{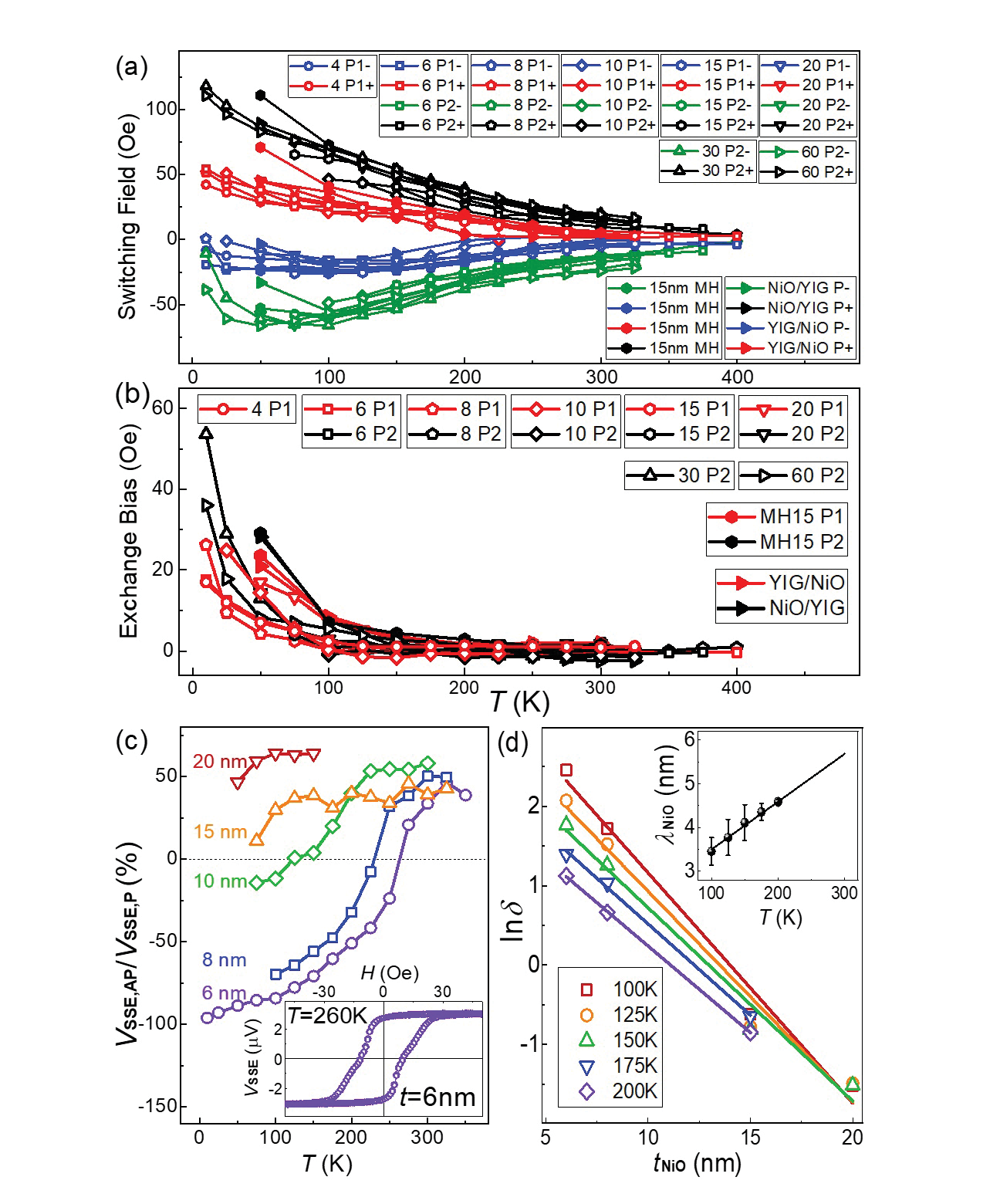}
\caption{\label{Fig4} \raggedright The dependence of (a) switching fields and (b) exchange bias of all the IMJs and the control samples on temperatures. (c) $V_\mathrm{SSE,AP}$/$V_\mathrm{SSE,P}$ ratios for the IMJs. Inset shows the field dependence of $V_\mathrm{SSE}$ for the IMJ with 6 nm NiO spacer and sizeable on-off ratio at 260 K. (d) Thickness dependence of ln$\delta$ at medium temperatures. Insets show derived $\lambda_\mathrm{NiO}$ as function of $T$.}
\end{figure}

\section{(6) Exchange Bias Dependence of SSE of IMJs}
We have also changed exchange bias direction of the same IMJ with 8 nm NiO by field cooling technique. In this case, we first elevated temperature to 400 K (the highest temperature of our PPMS system) and then applied 5 T field along different directions (along the Pt strip or in-plane vertical to the Pt stripe or normal to stacks) and then cooled the device down to 10 K with the field of 5 T maintained. Finally, the high field was reduced to 0 in oscillating mode to minimize remanence field of magnet. After all the above procedures, we began SSE measurement with the Pt stripe along the $y$ axis and field applied along the $x$ axis. At 10 K, only the case with cooling field in-plane vertical to the stripe shows remarkable exchange bias effect while the other two cases show small or even negligible exchange bias effect. This means we have successfully changed the exchange bias directions by the above procedures. However, as shown in Fig.7, all the three cases show nearly the same saturation $V_\mathrm{SSE}$ at all the temperatures. Furthermore, we can see from the data between 150 K and 300 K that magnon valve ratio was also independent on exchange bias directions. It indicates different exchange coupling directions at the interfaces would not deteriorate transfer efficiency of magnon current across the NiO/YIG interfaces, which is luckily benefit for applications. Independence of spin torque transfer on direction of exchange bias was very recently reported in metallic system~\cite{ref29}. Our data show this independence was also solidly reproduced for magnon transfer.

\begin{figure}[thb!]
\captionsetup{labelformat=default,labelsep=space} 
\includegraphics[width=8cm]{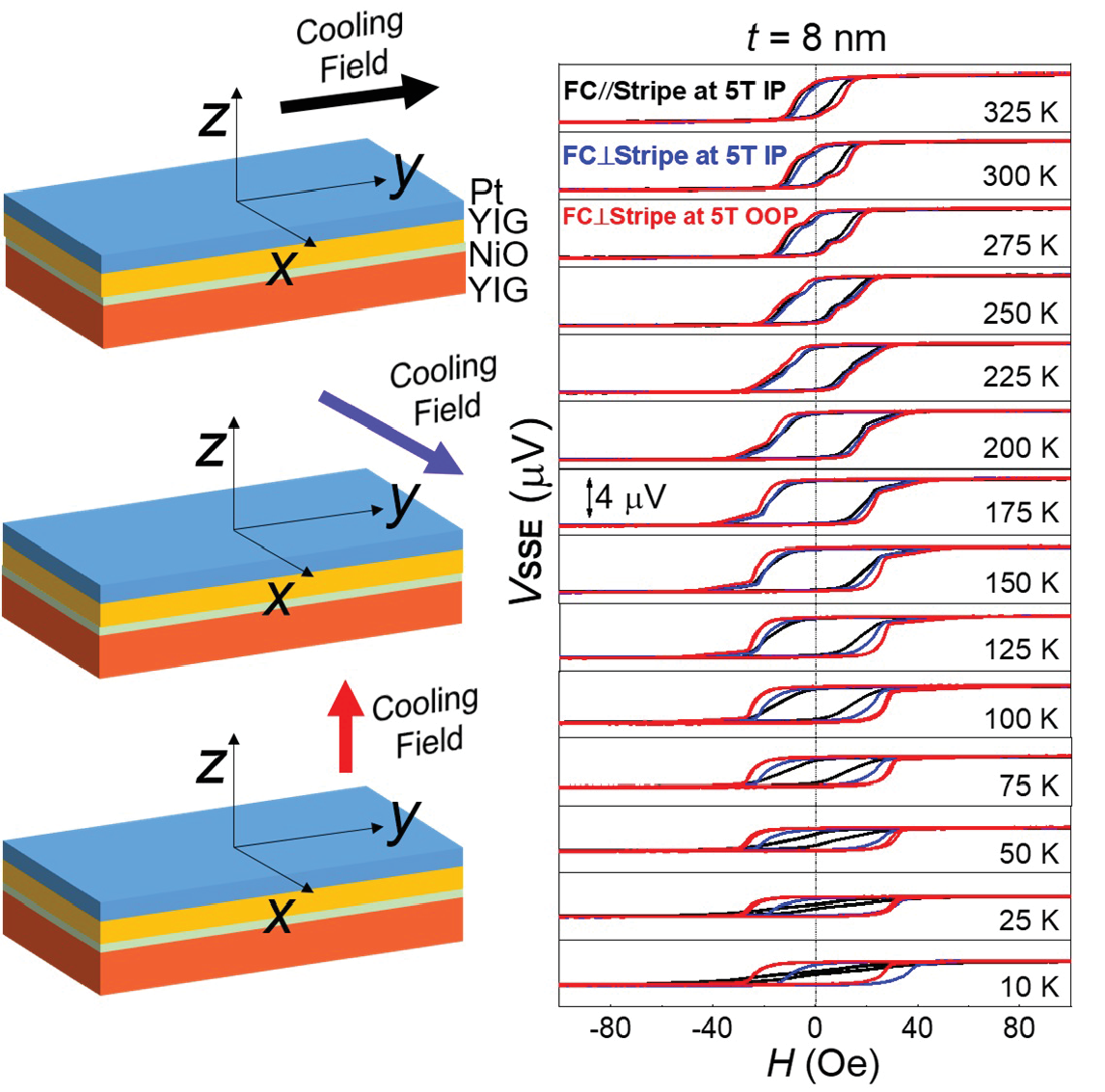}
\caption{\label{Fig4} \raggedright Independence of magnon transfer on directions of exchange bias. Three field cooling geometries are also shown as insets. Colors of the $V_\mathrm{SSE}$ vs. $H$ curves are the same with the cooling fields in the left insets. FC, IP and OOP denote cooling field, in-plane and out-of-plane, respectively.}
\end{figure}

\section{Conclusion}
In conclusion, fully electric-insulating and merely magnon-conductive magnon valves have been demonstrated by magnetic insulator YIG/antiferromagnetic insulator NiO/magnetic insulator YIG IMJs in which output spin current in Pt detector can be regulated with an high on-off ratio between P and AP states of the two YIG layers near room temperatures. The magnon current is dominated by the bottom (top) YIG layer at low (high) temperature regions. The transition temperature depends on $t_\mathrm{NiO}$. The magnon decay length in NiO is about several nanometers. Magnon transfer efficiency is independent on exchange bias directions. Most importantly, similar to the fundamental role played by magnetic tunnel junction (MTJ) in spintronics, the IMJ can also provide a basic building block for magnonics and oxide spintronics. Pure magnonic devices/circuits based on IMJs can be constructed in an ideally insulating system with magnon being the only angular momentum carrier. In these devices/circuits, information processing and transport can be accomplished only by magnons without mobile electrons and ultralow energy consumption and more versatile functions such as non-Boolean logics utilizing magnon phase coherence, magnetic memory based on insulators, magnon diode, transistors, waveguide and switches with large on-off ratios can be expected.
\section{Acknowledgements}
We gratefully thank Prof. S. Zhang in University of Arizona for enlightening discussions. This work was supported by the National Key Research and Development Program of China [MOST, Grants No. 2017YFA0206200], the National Natural Science Foundation of China [NSFC, Grants No.11434014, No.51620105004, and No.11674373], and partially supported by the Strategic Priority Research Program (B) [Grant No. XDB07030200], the International Partnership Program (Grant No.112111KYSB20170090), and the Key Research Program of Frontier Sciences (Grant No. QYZDJ-SSWSLH016) of the Chinese Academy of Sciences (CAS).






%


\end{document}